\documentclass{webofc}
\usepackage[varg]{txfonts}

\newcommand{\xpom}{x_\mathbb{P}}

\newcommand{\rt}{{\mathbf{r}_\perp}}

\newcommand{\bt}{{\mathbf{b}_\perp}}

\newcommand{\Deltat}{{\boldsymbol{\Delta}_\perp}}

\newcommand{\jpsim}{\mathrm{J}/\psi}

\begin{document}
\title{Probing nuclear structure at the Electron-Ion Collider and in ultra-peripheral nuclear collisions}

\author{ \firstname{Heikki} \lastname{M\"antysaari}\inst{1,2}
        \and
        \firstname{Farid} \lastname{Salazar}\inst{3,4}
        \and
        \firstname{Bj\"orn} \lastname{Schenke}\inst{5}
        \and
        \firstname{Chun} \lastname{Shen}\inst{6,7}
        \and
        \firstname{Wenbin} \lastname{Zhao}\inst{3,6}\fnsep\thanks{\email{WenbinZhao@lbl.gov}} 
}

\institute{
           Department of Physics, University of Jyv\"askyl\"a, P.O. Box 35, 40014 University of Jyv\"askyl\"a, Finland
\and       Helsinki Institute of Physics, P.O. Box                          64, 00014 University of Helsinki, Finland
\and       Nuclear Science Division, Lawrence Berkeley National             Laboratory, Berkeley, California 94720, USA
\and       Mani L. Bhaumik Institute for Theoretical Physics,               University of California, Los Angeles, California 90095, USA
\and
           Physics Department, Brookhaven National Laboratory, Upton, NY 11973, USA
\and        
           Department of Physics and Astronomy, Wayne State University, Detroit, Michigan 48201, USA
\and
           RIKEN BNL Research Center, Brookhaven National Laboratory, Upton, NY 11973, USA
          }

\abstract{%
Within the Color Glass Condensate framework, we demonstrate that exclusive vector meson production at high energy is sensitive to the geometric deformation of the target nucleus and subnucleon scale fluctuations.
Deformation of the nucleus enhances the incoherent cross section in the small $|t|$ region. Subnucleon scale fluctuations increase the incoherent cross section in the large $|t|$ region.  In ultra-peripheral collisions (UPCs),  larger deformation leads to a wider distribution of the  minimal impact parameter $B_{min}$ required to produce a UPC. This, together with larger incoherent cross sections for larger deformation, results in smaller extracted radii. Our results demonstrate great potential for future studies of nuclear structure in UPCs and electron-ion collisions.

}
\maketitle
\section{Introduction}
\label{sec:intro}
Understanding the geometric structure of nuclei is of fundamental interest. Exclusive vector meson (e.g.~$\mathrm{J}/\psi$) production in electron-ion collisions is a clean and powerful process to probe nuclei at small longitudinal momentum fraction $x$ for several reasons.
First, a large rapidity gap allows for the clean identification of an exclusive vector meson production process. Further, its cross section is approximately proportional to the squared gluon distribution~\cite{Ryskin:1992ui} (at leading order), providing increased sensitivity over inclusive processes. Finally, only in such exclusive scattering is it possible to determine the total momentum transfer to the target hadron or nucleus, which is the Fourier conjugate to the impact parameter and provides access to the target geometry.

Before facilities like the Electron-Ion Collider (EIC), LHeC/FCC-he, and EicC \cite{AbdulKhalek:2021gbh, LHeC:2020van,Anderle:2021wcy} are realized, ultra-peripheral collisions (UPCs) of heavy ions offer a great opportunity to explore nuclear structure with beams of quasi-real photons~\cite{Bertulani:2005ru,Klein:2019qfb}. In UPCs, the photon-nucleus ($\gamma+A$) interactions involving photons emitted from one of the colliding nuclei are dominant. Recently, STAR collaboration have revealed significant ${\rm cos} (2\Delta\Phi)$ modulations in exclusive $\rho$ meson photoproduction in UPCs~\cite{STAR:2022wfe} with $\Delta\Phi$ the angle between the produced $\rho$ meson transverse momentum $\mathbf{q}_\perp$ and its decay product pions' relative transverse momentum. This modulation originates from the fact that both nuclei can act as photon sources and as such there are two production channels whose interference need to be taken into account.
STAR also extracts the strong-interaction nuclear radii of ${\rm ^{197} Au }$ and ${\rm ^{238} U }$.   

In this work, we perform calculations of exclusive vector meson production in electron + uranium collisions and ultra-peripheral U+U collisions with different deformation parameters within the Color Glass Condensate (CGC) framework~\cite{Mantysaari:2022ffw,Mantysaari:2023qsq,Mantysaari:2023prg}. 

\section{Model}
\label{sec:model}
The exclusive production of a vector meson, $\gamma^* + p/A \to V + p/A $, provides insight into the small-$x$ structure of the target nuclei. The coherent cross section, corresponding to the events where the target remains intact, is obtained
by averaging the amplitude over the target color charge configurations before squaring.  The incoherent vector meson production, for which the final state of the target is different from its initial state, has the form of a variance,
\begin{equation}
\label{eq:coherent}
     \frac{\mathrm{d} \sigma^{\gamma^* + A \to V + A}_{\rm Coh.}}{\mathrm{d} |t|}  = \frac{1}{16\pi} \left|\left\langle \mathcal{A} \right\rangle_\Omega\right|^2, \   \ 
     \frac{\mathrm{d} \sigma^{\gamma^* + A \to V + A^*}_{\rm Incoh.}}{\mathrm{d} \vert t\vert}  = \frac{1}{16\pi} \left[
     \left\langle \left|\mathcal{A}\right|^2\right \rangle_\Omega \right. 
     - \left. \left|\left\langle \mathcal{A}\right\rangle_\Omega\right|^2 \right]\,.
\end{equation}
Here $\langle...\rangle_\Omega$ correspond to an average over the target configurations $\Omega$. 
The scattering amplitude for exclusive vector meson production in the dipole picture is
\begin{equation}
\label{eq:jpsi_amp}
    \mathcal{A} = 2i\int d^{2}{\rt} d^{2}{\bt}  \frac{d{z}}{4\pi} e^{-i \left[\bt - \left(1/2-z\right)\rt\right]\cdot \Deltat}  \times [\Psi_V^* \Psi_\gamma](Q^2,\rt,z) N_\Omega(\rt,\bt,\xpom).
\end{equation}
Here, $\rt$ is the transverse size of the $q\bar q$ dipole,  $\bt$ is the impact parameter measured relative to the target center, and $Q^2$ is the photon virtuality. The fraction of the large photon plus-momentum carried by the quark is given by $z$, $\xpom$ is the fraction of the target longitudinal momentum transferred to the meson in the frame where the target has a large momentum, and $\Deltat$ is the transverse momentum transfer, with $-t \approx {\boldsymbol \Delta}_\perp^2$. The $\gamma^* \to q\bar q$ splitting is described by the virtual photon light front wave function $\Psi_\gamma$. The $\Psi_V$ is the vector meson wave function, for which we use the Boosted Gaussian parametrization from~\cite{Kowalski:2006hc}.

In UPCs, there are two indistinguishable contributions to vector meson production. Either of the colliding nuclei can act as a source of linearly polarized photons and the other being the target. We take into account the quantum mechanical interference of these two processes.
We also implement the non-zero transverse momentum of the incoming photons, which significantly impacts the transverse momentum spectra of the produced vector mesons \cite{Mantysaari:2022sux}. This effect is particularly important near the diffractive minima of the coherent cross-section.

\begin{figure*}[t]
\begin{center} \includegraphics[scale=0.3]{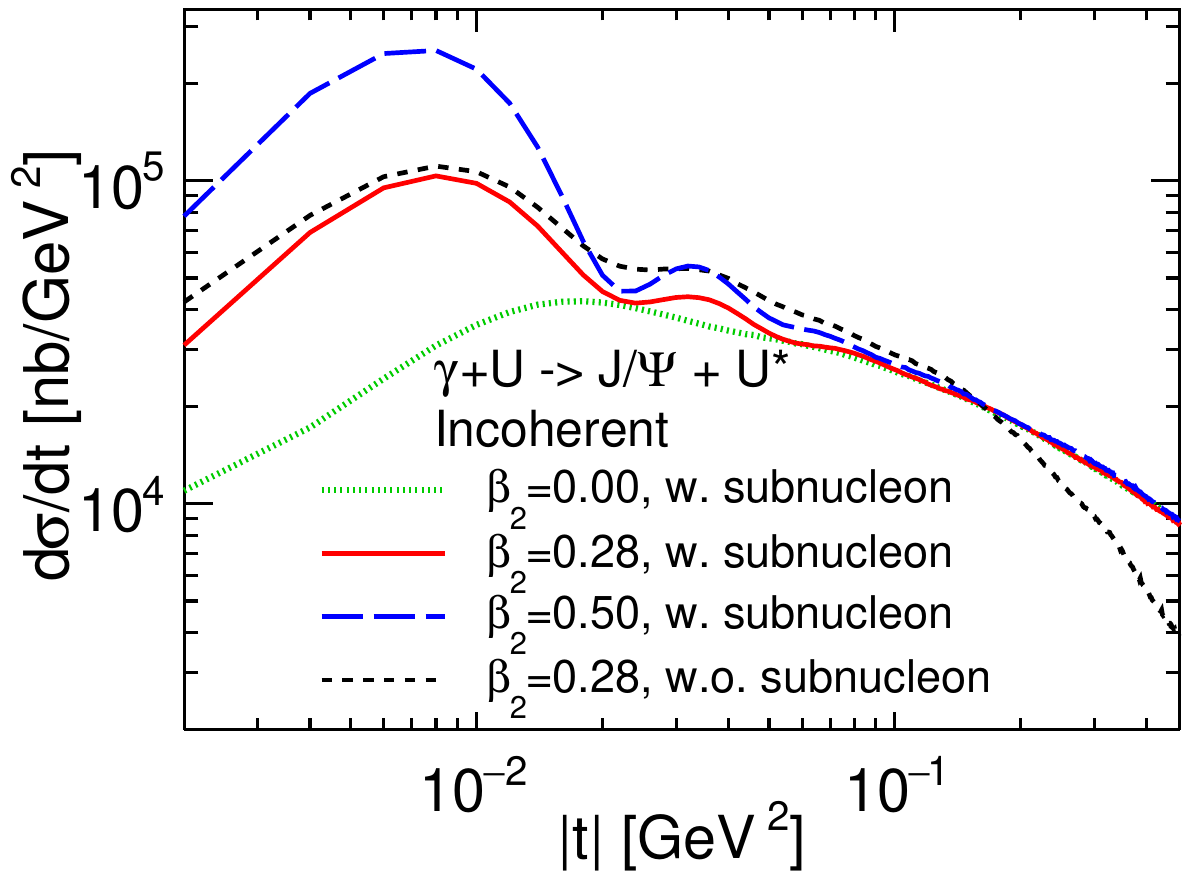}
  \end{center}  \vspace{-0.25cm}
  \caption{(Color online) The incoherent $\mathrm{J}/\psi$ photoproduction cross sections at $x_p=1.7 \times 10^{-3}$ in e+U collisions for different $\beta_2$ values and with or without subnucleonic structure fluctuations. }
  \label{fig:eU}
  \vspace{-0.25cm}
\end{figure*}

\begin{figure*}[t]
  \includegraphics[scale=0.3]{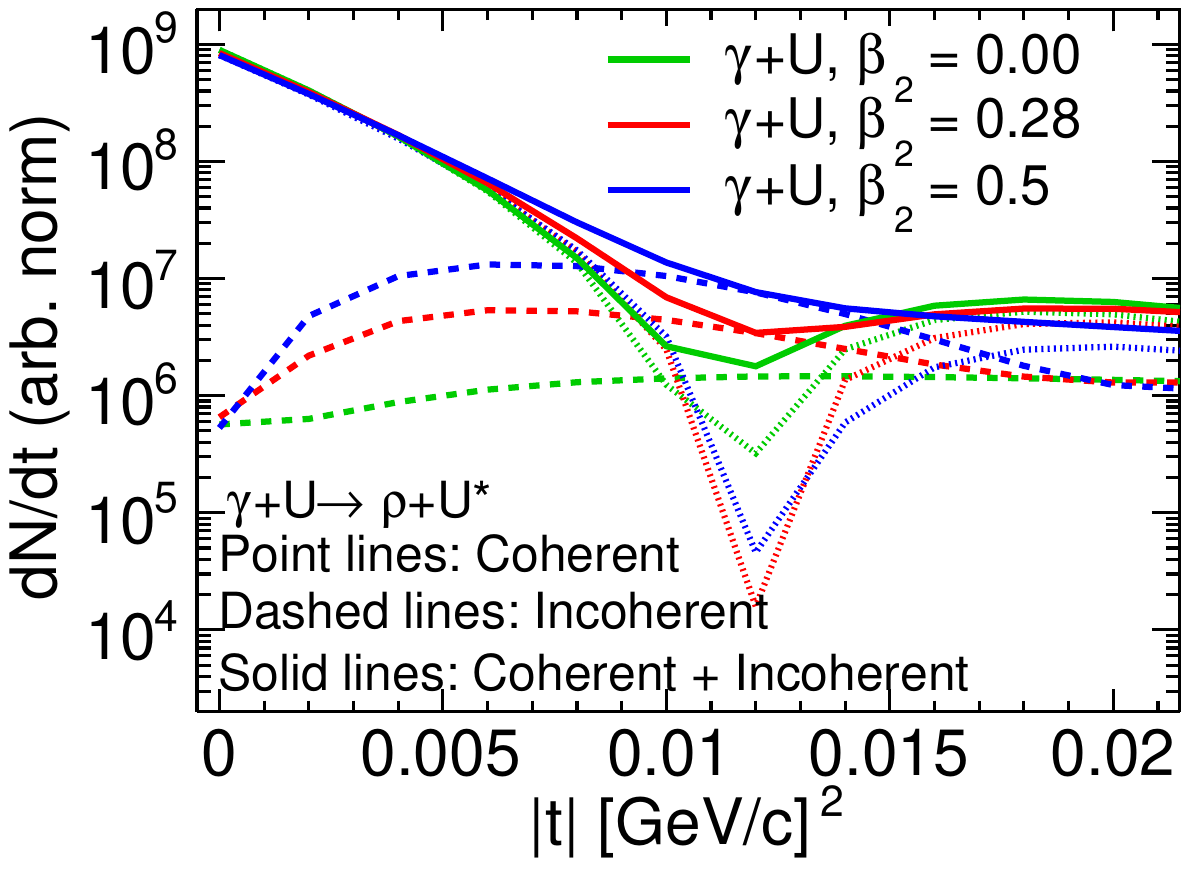}
  \includegraphics[scale=0.3]{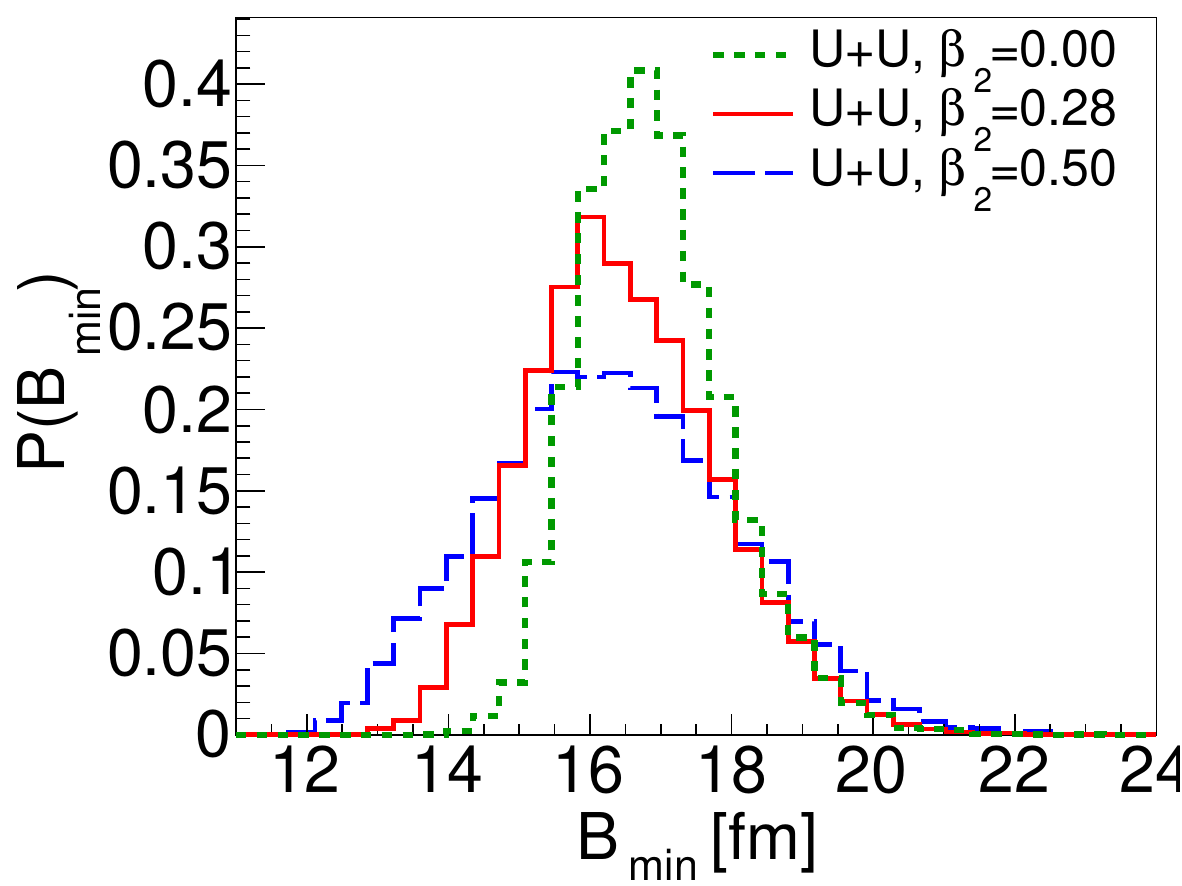}
  \caption{(Color online) Left panel: The coherent, incoherent, and total  $\rho$ photoproduction cross sections  in e+U collisions for different $\beta_2$ values. Right panel: The  distributions of minimal impact parameters $B_{\rm min}$ for UPCs between the two uranium nuclei at 193 GeV for different $\beta_2$ values. }
  \label{fig:eU_pB}
  \vspace{-0.5cm}
\end{figure*}

\section{Results}
Figure \ref{fig:eU} shows the effects of quadrupole ($\beta_2$) deformation and of subnucleonic structure fluctuations on the incoherent $\gamma + \mathrm{U} \to \jpsim + \mathrm{U}^{(*)}$ cross sections. 
As the quadrupole deformation increases, there is a noticeable rise in the incoherent cross section, particularly at small $\vert t\vert < 0.02~{\rm GeV^2}$.
This enhancement can be attributed to the random orientations of the deformed nucleus in the laboratory frame, leading to larger density fluctuations  of configurations projected onto the transverse plane. Consequently, larger event-by-event fluctuations in scattering amplitude occur. Similar effects are reported for octupole ($\beta_3$) and hexadecapole ($\beta_4$) deformations in \cite{Mantysaari:2023qsq}.
Comparing the dashed black line, representing the case without subnucleonic fluctuations, to the red line with subnucleonic fluctuations, a distinct trend emerges at $\vert t\vert > 0.25~ {\rm GeV^2}$. The faster drop in the black line demonstrates that subnucleonic fluctuations alter the slope and magnitude of the incoherent cross section at large $\vert t\vert$. 
The figure shows the potential of such a measurement for imaging nuclear properties, ranging from deformations to subnucleonic fluctuations across orders of magnitude in momentum transfer.

The nuclear deformation was found to play a significant role in imaging the nucleus' averaged radius in e+A collisions and UPCs \cite{Mantysaari:2023prg}. The left panel of Figure \ref{fig:eU_pB} shows the cross sections for $\rho$ production in $\gamma+ \mathrm{U} \to \rho + \mathrm{U}^{(*)}$ at $Q^2=0.0 {\rm GeV^2}$ processes, with varying $\beta_2$ values. The increase in the degree of deformation leads to an enhancement of the incoherent cross section. This enhancement produces a flatter total $dN/d|t|$ distribution. Since only the total cross section is accessible experimentally, we extract the uranium radius by fitting the total $\mathrm{d}N/\mathrm{d}|t|$. It is found that the slope of the total cross section is flatter for U with a large deformation, which would result in a smaller extracted radius.
The right panel of Figure \ref{fig:eU_pB} shows the minimal impact parameters $B_\mathrm{min}$ distributions for UPCs between the two uranium nuclei for different $\beta_2$ values. Larger $\beta_2$ values lead to wider $B_\mathrm{min}$ distributions. Configurations that allow for
small $B_\mathrm{min}$, with greater interference and larger photon transverse momentum, result in flatter spectra and enter with a larger weight. These two effects result in the larger effective radii $R$ for smaller deformations of the nucleus. For details, please refer to \cite{Mantysaari:2023prg}.

\section{Conclusions}
We have demonstrated that nuclear deformations can significantly modify the incoherent vector meson 
production cross section e+A collisions. 
Nuclear deformations and subnucleonic fluctuations on different length scales exhibit distinct effects in different momentum transfer regions. This result suggests that future EIC experiments can disentangle fluctuations at multiple length scales.
A larger deformation leads to smaller extracted radii because of the increased incoherent contribution and, in UPCs, because of how the deformation affects the minimal impact parameter $B_\mathrm{min}$  distribution. 
This study offers alternative methods to access nuclear properties. Combining traditional nuclear structure experiments, electron-ion, and ultra-peripheral collisions could provide unprecedented insights in the future.

\section*{Acknowledgments}
This material is based upon work supported by the U.S. Department of Energy, Office of Science, Office of Nuclear Physics, under DOE Contract No.~DE-SC0012704 (B.P.S.) and Award No.~DE-SC0021969 (C.S.), and within the framework of the Saturated Glue (SURGE) Topical Theory Collaboration.
C.S. acknowledges a DOE Office of Science Early Career Award. 
H.M. is supported by the Research Council of Finland, the Centre of Excellence in Quark Matter, and projects 338263 and 346567, and under the European Union’s Horizon 2020 research and innovation programme by the European Research Council (ERC, grant agreement No. ERC-2018-ADG-835105 YoctoLHC) and by the STRONG-2020 project (grant agreement No 824093)
F.S. and W.B.Z. are supported by the National Science Foundation (NSF) under grant number ACI-2004571 within the framework of the XSCAPE project of the JETSCAPE collaboration.
The content of this article does not reflect the official opinion of the European Union and responsibility for the information and views expressed therein lies entirely with the authors.
This research was done using resources provided by the Open Science Grid (OSG)~\cite{Pordes:2007zzb, Sfiligoi:2009cct} \#2030508.

\bibliography{references}

\begin{thebibliography}{14}

\bibitem{Ryskin:1992ui}
M.G. Ryskin, Z. Phys. C \textbf{57}, 89 (1993)

\bibitem{AbdulKhalek:2021gbh}
R.~Abdul~Khalek et~al., Nucl. Phys. A \textbf{1026}, 122447 (2022),
  \texttt{2103.05419}

\bibitem{LHeC:2020van}
P.~Agostini et~al. (LHeC, FCC-he Study Group), J. Phys. G \textbf{48}, 110501
  (2021), \texttt{2007.14491}

\bibitem{Anderle:2021wcy}
D.P. Anderle et~al., Front. Phys. (Beijing) \textbf{16}, 64701 (2021),
  \texttt{2102.09222}

\bibitem{Bertulani:2005ru}
C.A. Bertulani, S.R. Klein, J.~Nystrand, Ann. Rev. Nucl. Part. Sci.
  \textbf{55}, 271 (2005), \texttt{nucl-ex/0502005}

\bibitem{Klein:2019qfb}
S.R. Klein, H.~M\"antysaari, Nature Rev. Phys. \textbf{1}, 662 (2019),
  \texttt{1910.10858}

\bibitem{STAR:2022wfe}
M.~Abdallah et~al. (STAR), Sci. Adv. \textbf{9}, eabq3903 (2023),
  \texttt{2204.01625}

\bibitem{Mantysaari:2022ffw}
H.~M\"antysaari, B.~Schenke, C.~Shen, W.~Zhao, Phys. Lett. B \textbf{833},
  137348 (2022), \texttt{2202.01998}

\bibitem{Mantysaari:2023qsq}
H.~M\"antysaari, B.~Schenke, C.~Shen, W.~Zhao, Phys. Rev. Lett. \textbf{131},
  062301 (2023), \texttt{2303.04866}

\bibitem{Mantysaari:2023prg}
H.~M\"antysaari, F.~Salazar, B.~Schenke, C.~Shen, W.~Zhao (2023),
  \texttt{2310.15300}

\bibitem{Kowalski:2006hc}
H.~Kowalski, L.~Motyka, G.~Watt, Phys. Rev. D \textbf{74}, 074016 (2006),
  \texttt{hep-ph/0606272}

\bibitem{Mantysaari:2022sux}
H.~M\"antysaari, F.~Salazar, B.~Schenke, Phys. Rev. D \textbf{106}, 074019
  (2022), \texttt{2207.03712}

\bibitem{Pordes:2007zzb}
R.~Pordes et~al., J. Phys. Conf. Ser. \textbf{78}, 012057 (2007)

\bibitem{Sfiligoi:2009cct}
I.~Sfiligoi, D.C. Bradley, B.~Holzman, P.~Mhashilkar, S.~Padhi, F.~Wurthwrin,
  WRI World Congress \textbf{2}, 428 (2009)

\end{thebibliography}

\end{document}